\documentclass[a4paper,12pt,oneside]{article}
\usepackage[koi8-r]{inputenc}
\usepackage[english,russian]{babel}
\usepackage{graphicx}
\begin{document}

\title{
The Structure of Shock Waves in the Case of Accretion onto Low-Mass Young Stars
 }

\author{Lamzin S.A.}

\date{\it Sternberg Astronomical Institute, 
Universitetskii prospekt 13, Moscow, 119899 Russia\\
\small Published in Astronomy Reports, Vol. 42, No 3, 1998, pp. 322-335.}

\maketitle

\section*{Abstract}

\bigskip

  A physical analysis and mathematical formulation of the structure 
of an accretion shock wave in the case of young stars are given. 
Some results are presented, and the dependence of the structure of
the shock on various parameters is investigated. In the general case, 
the relative intensities and profiles of lines in the spectra of 
T Tauri stars should depend on the velocity and density of the 
infalling gas, as well as on the accretion geometry. The calculation 
results may serve as a base for elucidating the nature of the activity 
of young stars.

\bigskip
\bigskip

\section{Introduction}

   T Tauri stars are young ($t \le 10^7$ years), low-mass 
($M \le 2M_\odot$) stars at the stage of gravitational 
contraction toward the main sequence. From the observational 
point of view, they are late-type subgiants with emission lines 
of hydrogen and Ca\,II, as well as of some other atoms and ions, 
mostly with relatively low ionization potentials. T Tauri stars 
are also characterized by excess continuum emission at short visible 
wavelengths and in the infrared compared to main-sequence stars of 
the same spectral types (see the reviews [1, 2]). The variety and 
relative intensities of their emission-line spectra, as well as 
their continuum excesses, correlate with the equivalent width of 
the $H_\alpha$ line; thus, we may consider this line to be an 
indicator of the activity of low-mass young stars. It is accepted 
to call stars with $W_{H_\alpha} > 5-10$ {\AA} classical T Tauri 
stars (CTTSs), and precisely such stars will be the subject of our 
studies below.

   Infrared and radio investigations of CTTSs testify that these 
stars are surrounded by disklike gas-dust envelopes. IUE observations 
of CTTSs show the presence of lines of highly ionized atoms, up to 
N\,V; the luminosity in these lines is several orders of magnitude 
higher than for the corresponding solar lines [3]. However, at X-ray 
energies, between 0.5 and 4.5 keV, CTTSs have luminosities $\le10^{31}$ 
erg/s, i.e., 2-3 orders of magnitude lower than expected by analogy 
with the Sun based on the observed intensities of, e.g., the C\,IV 
$\lambda$1550 line. The absence of forbidden Fe\,X $\lambda$6375 and 
Fe\,XIV $\lambda$5303 coronal lines in the spectra of CTTSs suggests 
that this is not associated with the absorption of X-rays in the extended 
circumstellar envelopes, but rather is due to a real difference of 
the observed $L_X/L_{C\,IV}$ ratio from the solar ratio. The comparatively 
hard spectrum $(T \ge 3\times10^6$~K) and variability of the observed 
X-ray emission suggest that it is associated with flare activity [2, 4].

  There is not yet any consensus about the nature of the regions in which 
the CTTS emission-line spectrum forms, nor about the origin of the rather 
strong outflows of matter from the vicinitiy of CTTSs in the form of 
so-called CO flows and jets. Up until the early 1980s, it was thought 
that the activity of CTTSs was the result of powerful chromospheres 
and/or coronae, to some extent similar to the corresponding solar structures, 
and that the reason for the existence of hot layers above the photosphere 
and of the stellar wind was the dissipation of mechanical energy generated 
by the stellar convective zone [5-9]. Later, it was proposed that the 
emission-line spectrum was associated with heating of material from the disklike
circumstellar envelope falling onto the young star. If the inner edge of the disk 
is in contact with the stellar surface, the heating should take place in a boundary 
layer, where the disk material is decelerated from the Keplerian velocity 
to the stellar rotation velocity [10, 11]. However, if the young star has 
a fairly powerful dipolar magnetic field, then, by analogy with X-ray pulsars, 
it may be that the inner disk boundary does not reach the stellar surface, 
and that the disk material one way or another penetrates the magnetosphere 
and slides down along lines of force. In this case, it is natural to associate 
the region of formation of the emission-line spectrum with the shock wave 
that should arise during the collision of the infalling gas with the stellar 
surface [12-14]. In accretion models, the observed outflow of matter is 
interpreted in terms of a magnetohydrodynamic wind from the disk surface or 
directly from the stellar magnetosphere [2, 12, 15].

In principle, investigations of the profile shapes and relative intensities 
of CTTS emission lines could elucidate the nature of the hot regions in young 
stars. However, numerous attempts to use lines of atoms and ions with low 
ionization potentials (e.g., the hydrogen Balmer lines) in such studies have 
not yet been successful: similar profiles are obtained for quite different 
assumptions about the geometry, velocity field, and physical conditions in 
the emission-line formation region [1,2,16,17]. Moreover, high-angular-resolution 
spectral observations have shown that the profiles of these lines are formed 
both near the stellar surface and in the extended circumstellar envelope [18, 19].

Analyses of the line emission of ions with ionization potentials above 20~eV 
are much more promising in this connection [20]. At the present, the following 
is known about the profiles of these lines.

(1) The He\,II $\lambda$4686 lines in CTTS spectra have equivalent widths 
of less than 1 \AA, and it became possible to investigate their profiles 
with sufficiently high signal-to-noise ratio only recently [21-23]. In all 
cases (nine stars), the profile of this line is asymmetric and redshifted 
by a few dozen km\,s$^{-1}$.

(2) In [24], Gomez de Castro et al. analyze the profiles of the resonant 
lines of the Si\,III, Si\,IV, C\,III, and C\,IV ions using archival 
high-resolution IUE spectra of four CTTSs (T Tau, RU Lup, DR Tau, TW Hya) 
obtained in the 1980s. The profiles of lines that could be reliably 
distinguished against the noise background were asymmetric and redshifted. 
In the case of the C\,IV $\lambda$1550 doublet lines, two peaks of comparable 
intensities were reliably identified, with their maxima shifted by 
approximately $+50$ and $+250$ km\,s$^{-1}$ with respect to the laboratory wavelength.

(3) In the spectra of many CTTSs, there are He\,I emission lines that 
are about an order of magnitude more intense than the He\,II $\lambda$4686 
line. According to [1], the He\,I line profiles have a broad 
(FWHM $\sim 150$ km\,s$^{-1}$) peak that is symmetric with respect to the 
laboratory wavelength. In a number of cases, a narrower (FWHM $\le 50$ km\,s$^{-1})$ 
central component is superimposed. On the other hand, in high-resolution 
spectrograms with good signal-to-noise ratio, some stars have a broad 
absorption feature in the red wing of the $\lambda$5876 triplet; 
in RY Tau and BM And, the entire $\lambda$5876 line is in absorption, 
not emission [25].

  In [24, 25], it was proposed that the above features of the lines of 
helium and ions with ionization potentials above 20 eV can be understood 
if these lines form in a shock wave near the surface of the young star. 
Note that the observations currently available do not exclude the possibility 
that CTTSs have a comparatively powerful chromosphere and/or corona, which may 
contribute to the line emission of ions with high degrees of ionization. 
Therefore, studies of the emission of an accretion shock wave should also 
elucidate whether or not powerful chromospheres are present in low-mass young stars.

After calculating the shock-wave structure, we can find the intensities 
of specially selected "signal"{} lines; comparisons with observations enable 
us to determine the velocity and density of the accreted gas, and also the total 
accretion rate and the area of the stellar surface on which the material falls. 
To determine the "signal"{} line profiles, it is necessary to appropriately sum 
the emission from different points in the shock wave, so that analysis of the 
observed profiles and, in particular, their variation with the axial stellar 
rotation should yield information about the geometry of the accretion region.

The structure of radiative shocks has been investigated in a number of studies 
in connection with various problems in the physics of the interstellar medium 
[26, 27] and processes in the envelopes of nonstationary stars [28]. Therefore, 
we should explain the peculiarities of the study of such shocks in the case of 
CTTSs, and why we cannot apply already available results of gas-dynamical 
calculations to the interpretation of the spectra of young stars. The difference 
from the case of the interstellar medium is due to the fact that we consider 
a shock in a rather dense gas in which the velocity of the stream flowing into 
the front is $150-450$ km\,s$^{-1}$. The much higher density means, for instance, 
that the forbidden lines of O\,I, O\,II, N\,II, etc., cannot play an important 
role in cooling the gas. Moreover, in an interstellar medium with velocities 
higher than 100 km\,s$^{-1}$, postshock radiative cooling does not play a 
significant role, and the gas relaxes adiabatically [26]. On the other hand, 
shock waves in the envelopes of nonstationary stars are characterized by higher 
densities and velocities for the stream flowing into the front that are a factor 
of 2-3 lower [28]. Thus, the situation of interest to us is, in some sense, 
intermediate, and has not previously been investigated to our knowledge.

Before addressing the problem of the structure of an accretion shock wave 
in the case of a low-mass young star, we made some physical assumptions 
reflecting the specific character of the situation investigated, which made 
it possible to considerably simplify the general system of radiative gas-dynamical 
equations. Therefore, the main goal of this work is to show that the results of 
our numerical calculations do not contradict our initial assumptions, and can 
serve as a basis for the quantitative interpretation of the spectra of young 
stars. In the last section, we will present data on the structure of accretion 
shocks and the calculated intensities of some observable resonant lines.

\section{Formulation of the problem}

Consider a young star with mass $M_*$ and radius $R_*$ with a large-scale 
dipolar magnetic field surrounded by an accretion disk. Suppose that the 
stellar magnetic field stops the accretion disk at some distance $R_1>R_*$; 
the value of $R_1$ depends on the accretion rate $\dot{M}_a$ and the intensity 
of the stellar magnetic field [12-14]. The disk material is frozen in the magnetic 
lines of force and slides along them under the action of the force of gravity, 
acquiring near the stellar surface a radial velocity [29]
$$V_0\simeq\sqrt{\frac{2GM_*}{R_*}\left(1-\frac{R_*}{R_1}\right)}.$$
(Here, we assume that the angular velocity of the stellar rotation is much lower 
than the Keplerian velocity $(\omega_K=(GM_*)^{1/2}R_*^{-3/2}).$ According to the 
data of [30] and [25], for the stars RU Lup and RY Tau, $V_0\simeq170$ and 
400~km\,s$^{-1}$, respectively. Therefore, we will be interested in velocities 
for the infalling gas from 150 to 450 km\,s$^{-1}$.

During the collision of the gas with the stellar surface, a shock wave should 
arise. Let us assume that this shock has the following qualitative structure 
[30]. The gas compressed in the shock front is strongly heated, then gradually 
decelerates and cools to a temperature of the order of the stellar effective 
temperature, smoothly passing into the stellar atmosphere, which is in 
hydrostatic equilibrium. The cooling takes place via volume losses to emission 
in resonant lines of the ions of the most abundant elements; the largest fraction 
of the energy is in photons with energies above 13.6 eV. Half of the photons from 
the cooling zone move away from the star and are absorbed by the preshock matter, 
creating a heating and ionization zone that is transparent in the continuum at 
$\lambda > 912$~\AA. The other half of the photons are absorbed in the transition 
region between the radiative-cooling zone and the stellar atmosphere, and are then 
reemitted in the continuum. Only a small fraction of these photons goes into the 
Lyman continuum, so that the emission of the transition region and, consequently, 
its structure, virtually do not influence the structure of the higher-lying regions.

Within the framework of these assumptions, we estimated in [30] the 
number density $N_0$ (cm$^{-3})$ of the accreted gas far upstream from 
the front: it turned out that situations in which $10< \log N_0 < 13$ 
are of the most interest. For this interval of densities, we will also 
assume that the typical dimensions of the postshock radiative-cooling 
region and of the preshock H\,II region are smaller than the stellar 
radius, and are also smaller than the extent of the front perpendicular 
to the gas motion. Therefore, we will consider the shock wave to be 
plane-parallel and will solve the corresponding one-dimensional problem. 
Finally, we assume that the shock front is at rest with respect to the 
stellar surface, and that all the shock parameters depend only on the 
spatial coordinate. Thus, we consider below a solution of the one-dimensional, 
radiative, gas-dynamical equations describing the structure of an accretion 
shock wave that is stationary in Eulerian coordinates.

\section{System of radiative hydrodynamical equations}

   In a stationary, one-dimensional shock wave, all parameters depend 
only on the spatial coordinate z, and it is assumed that ${\left( 
\partial / \partial t \right)}_z=0$ everywhere in the gas-dynamical 
equations. We will also assume that a chaotic magnetic field is frozen 
in the infalling plasma, and that this field is much weaker than the 
large-scale stellar field channeling the flow of matter, but is still 
sufficiently intense to suppress the electron thermal conductivity. 
We are interested only in the field component perpendicular to the 
$z$ axis; we designate the intensity of this component $H.$

We will take the accreted matter to consist of a mixture of H, He, 
C, N, O, Ne, Mg, Si, S, and Fe, with the relative abundances
$$\zeta_a = \frac{N^a}{\sum\limits_a{N^a}}\equiv\frac{N^a}{N},$$
where $N^a$ and $N$ are the volume number densities of a given element 
[31] and the total number density of the nuclei of all elements, 
respectively. We denote $N^a_i$ and $x^a_i$ to be the volume and 
relative number densities of ions of the $i$th degree of ionization 
of element $a$ $(0 \le i \le Z_a,$ where $Z_a$ is the nuclear charge 
of this element), so that
$$N^a = \sum\limits_{i=0}^{Z_a}{N^a_i}, \qquad \sum\limits_{i=0}^{Z_a}x^a_i=1.$$
Let $N_e$ be the volume number density of free electrons. Then, 
designating $x_e = N_e/N$, we can write the equation of electric 
neutrality of the plasma
$$x_e = x^{\rm  H}_1+x^{\rm He}_1+2x^{\rm He}_2+3\times10^{-4},\eqno(1)$$ 
where the last term takes into account the single ionization of C\,I 
by lines of the hydrogen Lyman series in preshock H\,I regions. Note 
that, at this stage, we are interested solely in lines of ions with 
high ionization potentials (see above). Therefore, we will not 
investigate the structure of the H\,I regions in detail; we will show 
later, however, that these regions virtually do not affect the fluxes 
in the lines of interest to us.

In the absence of molecular hydrogen, the continuity equation is [32]
$$j \equiv NV = const = N_0V_0,\eqno(2)$$	
Here and below, all quantities with the subscript 0 refer to the preshock 
unperturbed flow, i.e., formally to $z = -\infty.$ We will assume that the 
gas temperatures for the various ions and atoms are the same and are equal 
to $T_i,$ but, generally speaking, are different from the temperature $T_e$ 
of the electron gas. We will express the gas density in terms of the mean 
molecular weight $\mu_i:$
$$\rho = \sum\limits_a{m_aN^a}\equiv \mu_im_{\rm H}N, \qquad \mu_i = \sum\limits_a\zeta_aA^a\simeq1.3,\eqno(3)$$ 
where $m_a$ and $A^a$ are the mass and atomic weights of an element. 
Since the pressure of an ideal gas is $P = NkT,$ the momentum conservation 
law in our case can be written
$$NkT_i + N_ekT_e + \mu_im_{\rm H}NV^2+\frac{H^2}{8\pi}=const\simeq N_0kT_{i0}+\mu_im_{\rm H}N_0V^2_0+\frac{H^2_0}{8\pi}, \eqno(4)$$
assuming that the gas is weakly ionized at $z =-\infty.$ Another algebraic 
equation describing the flux freezing of the small-scale chaotic magnetic 
field is [33]:
$$\frac{H}{N}=\frac{H_0}{N_0.}\eqno(5)$$ 

We will now consider the energy-conservation equations. In the two-temperature 
approximation adopted, separate equations must be written for the electrons and 
heavy particles, i.e., atoms and ions. For the heavy particles, the energy sink 
is the exchange of energy with electrons, and therefore the energy conservation 
law for them per heavy particle is [32]
$$V\frac{\rm d}{ \rm d z}\left(\frac{3}{2}kT_i\right)+VNkT_i\frac{\rm d}{ \rm d z}\left(\frac{1}{N}\right) = -\omega_{ei}N_e. \eqno(6a)$$ 
Here, we have used the stationarity condition, which allows us to replace 
$\frac{\rm d}{ \rm d t}$ with $V\left(\frac{\rm d}{ \rm d z}\right).$ Energy 
exchange with ions is much more efficient than with neutral atoms. Since, 
for $V_0 \sim 300$ km\,s$^{-1},$ the most abundant element -- hydrogen -- 
is appreciably ionized over nearly the entire flow region of interest to us, 
we will take into account in (6a) only interactions between electrons and protons. 
In this case [32],
$$\omega_{ei}\simeq\frac{3}{2}k\zeta_{\rm H}x^{\rm H}_1\frac{T_i-T_e}{T_e^{3/2}}\frac{\lambda_K}{250},$$
where $\lambda_K \simeq 9.4 + 1.5 \ln T_e - 0.5 \ln N_e $ is the so-called Coulomb logarithm.

For free electrons, the energy conservation law per heavy particle is [32]:
$$N\frac{\rm d}{ \rm d z}\left(\frac{3}{2}x_ekT_e\right)+VN_ekT_e\frac{\rm d}{ \rm d z}\left(\frac{1}{N}\right)=Q_{ph}-Q_{col}N_e+\omega_{ei}N_e. \eqno(7)$$ 
The quantity $Q_{ph}$ describes the heating of gas by photoionization; 
$Q_{col}$ denotes the sum of terms associated with energy losses to 
excitation $Q_{exc}$ and ionization $Q_{ion}$ via electron collisions, 
and with radiative cooling during recombinations $Q_{rec}.$ In other words, 
$$Q_{col}=Q_{exc}+Q_{ion}+Q_{rec}.$$ 
We will now write these terms in explicit form.

We assume that the cooling postshock gas is transparent to its own 
"cooling"{} radiation down to a temperature of $\sim10^4$~K, and that 
the heating and ionization of this gas by this radiation takes place 
{\it upstream} from the shock front and far {\it downstream} from the 
front, in the so-called transition region. We can write for the heating 
of gas by "external"{} radiation
$$Q_{ph}=\sum\limits_l\frac{F_l}{h\nu_l}
\sum\limits_a{\zeta_a}
\sum\limits_i{x_i^a}
\sum\limits_j{\sigma^a_{i,j,l}(h\nu_l-\chi^a_{i,j})}, \eqno(8)$$
where $F_l$, is the flux of radiation in the spectral line with 
frequency $\nu_l$; $\chi^a_{i,j}$ is the ionization potential of 
the $i$-th ion of element $a$ from shell $j$; $\sigma^a_{i,j,l}$ 
is the cross-section for photoionization of this ion by the line 
$l$; and the numerical value of $\nu_l;$ $\chi^a_{i,j}$ was 
calculated using formulas from [34]. (We also described the ionizing 
continuum radiation using certain fictitious lines -- see below). 
We considered photoionization from the inner K and L shells under 
the assumption that this is always accompanied by the Auger effect, 
excluding H-, He-, and Li-like ions. In equation (8), we took this 
into account by substituting $2\chi^a_{i,L}$ in place of 
$\chi^a_{i,K}.$ Since the flux of photons with energies above 
1~keV is small for the velocities $V_0$ of interest to us, we 
neglected the possibility of ionization from M shells.

For a normal abundance of heavy elements, energy losses due 
to collisional ionization are significant only for hydrogen and 
helium, so that
$$Q_{ion} = \zeta_{\rm H}x_0^{\rm H}q_0^{\rm H}\chi_0^{\rm H}+
                   \zeta_{\rm He}x_0^{\rm He}q_0^{\rm He}\chi_0^{\rm He}+
                   \zeta_{\rm He}x_1^{\rm He}q_1^{\rm He}\chi_1^{\rm He}, \eqno(9)$$
where $q^a_i$ is the coefficient for ionization by electron 
collisions. Energy losses via recombination were also taken into 
account only for hydrogen and helium. If an ensemble of free 
electrons loses an energy $\beta(T)kT$ to recombination, then (cf. [35])
$$Q_{rec} = kT\left(
\zeta_{\rm H}x_1^{\rm H}\beta^{\rm H}\alpha^{\rm H}_{1,0}+
\zeta_{\rm He}x_1^{\rm He}\beta^{\rm He}_1\alpha^{\rm He}_{1,0}+
\zeta_{\rm He}x_2^{\rm He}\beta^{\rm He}_2\alpha^{\rm He}_{2,1}
\right), \eqno(10)$$
where $\alpha^a_{i+1,i}$ is the coefficient for radiative 
recombination for an ion with charge $i+1$ to become an ion with 
charge $i.$ We assumed that the optical depth in the resonant lines 
of hydrogen and helium is fairly high, so that only recombinations 
to excited levels were taken into account in (10).

   We will now present an expression for the energy losses to 
excitation of discrete levels, with the subsequent emission of 
the corresponding photons. We considered electron collisional 
excitation only from the lowest term of the ground configuration, 
so that
$$Q_{exc} = \sum\limits_a{\zeta_a}
\sum\limits_i{x^a_i}
\sum\limits_k{h\nu_{1,k}\left(n_1C^{a,i}_{1,k}-n_kC^{a,i}_{k,1}\right)\epsilon_{1,k}}, \eqno(11)$$
where $n_k$ is the relative population of level $k,$ $C^{a,i}_{1,k}$ and $C^{a,i}_{k,1}$ are 
the coefficients of electron-collisional excitation and deexcitation, 
respectively, and $\epsilon_{1,k}$ is the probability of escape of a photon 
from the emission region, which we assume to be equal to 1. We took our list 
of the most intense lines (in total about 150) from [36], and the expressions 
for $C^{a,i}_{1,k}$ from the reviews [37].

Consider the ionization equilibrium equations. Let $I^{a}_i$ be the rate of 
photoionization of the $i$th ion of element $a$ from all its subshells per 
unit volume. Then, assuming $\sigma^a_{i,j,l}=0,$ if $h\nu_l < \chi^a_{i,j},$ 
we obtain
$$I^a_i = \sum\limits_l\frac{F_l}{h\nu_l}\sum\limits_j\sigma_{i,j,l}$$
for $i < Z_a$ and $I^a_i = 0$ for $i = Z_a;$ the subscript $j$ denotes 
the electron shells of the ion considered ($1 \le j \le j_{max}).$ $J^a_i$ 
is an analogous quantity characterizing the rate of photoionization only 
from those shells for which the removal of an electron is accompanied 
by the Auger effect.

   Generally speaking, the transition of an ion $i$ to the $i-1$ 
state may take place as a result of radiative, dielectronic, or 
three-particle recombination. In the situation considered here, 
nearly everywhere $\log N_e > 10.$ According to [38], at such densities, 
collisions with electrons disrupt highly excited ionic states, 
leading to a substantial decrease of the dielectronic recombination rate, 
since captures to high-lying levels play the major role in this process. 
For $\log N_e <15,$ the radiative recombination rates depend much more weakly 
on the density, since the contribution of high-lying states is much smaller 
in this case. Using the data in [39-41], we came to the following conclusions 
about the dielectronic recombination rate for the interval of densities 
and temperatures of interest to us: (1) for ions with charges smaller than six, 
it is considerably lower than the radiative recombination rate; (2) 
for hydrogen-and helium-like ions, density-dependent corrections do not 
exceed 20\%. Therefore, for the C\,II--C\,IV, N\,I--N\,V, O\,I--O\,VI, 
Ne\,I--Ne\,VI, Mg\,II--Mg\,VI, Si\,II--Si\,IV, S\,II--S\,VI, and Fe\,II--Fe\,VI 
ions, we set the dielectronic recombination coefficient to zero, while, 
for the remaining ions, we neglected the density dependence, as for 
the radiative recombination coefficient. In Section 5, we will present 
arguments indicating that this approximation is unlikely to introduce 
significant errors into the final result.

   Three-particle recombination plays an appreciable role only at rather 
high densities; the main contribution to this process comes from captures 
to high-lying levels. In particular, for hydrogen at a temperature of $10^4$~K 
and with $\log N_e = 14,$ the rate of three-particle recombination to the first 
level is approximately three orders of magnitude smaller than the total radiative 
recombination rate [32]. Interactions with ions and electrons in a plasma with 
such parameters limit the number of stable levels of the hydrogen atom to a value 
not greater than 100 [42]. Apparently, this means that, in the problem at hand, 
the rate of three-particle recombination for hydrogen nowhere exceeds 
$\sim10$\% of the radiative recombination rate; for this reason, we did not 
take it into account. Since the rate of three-particle recombination rapidly 
decreases with increasing nuclear charge [43], we likewise did not take it into 
account for the remaining ions.

We denote the radiative and dielectronic recombination coefficients 
$\alpha^a_{i,i-1}$ and $\kappa^a_{i,i-1},$ respectively. The number 
of recombinations involving the ith ion of element $a$ in unit volume 
per second can be written
$$R^a_{i,i-1}=N_e\left(\alpha^a_{i,i-1}+\kappa^a_{i,i-1}\right)$$
for $i = 1 - Z_a$ and
$$R^a_{i,i-1}=0$$
for $i = 0.$

  In terms of this notation, the ionization-equilibrium equations 
take the form (for simplicity we omit the subscript $a$):
$$V\frac{\rm d x_i}{\rm d z}=J_{i-2}x_{i-2}+
\left(q_{i-1}N_e+I_{i-1}-J_{i-1}\right)x_{i-1}-
\left(q_iN_e+I_i+R_{i,i-1}\right)x_i+R_{i+1,i}x_{i+1}. \eqno(12)$$
We calculated $\alpha(T),$ $\kappa(T),$ and $q(T)$ using 
formulas from [36]. For a number of ions, charge exchange reactions of the form
$$A^{+n}+{\rm H}^0 \rightarrow A^{+(n-1)}+{\rm H}^+,$$
$$A^{+n}+{\rm He}^0 \rightarrow A^{+(n-1)}+{\rm He}^+$$
play an important role in the ionization balance. Formally, we took this 
into account by adding the terms $N\zeta_{\rm  H}x^a_ix^{\rm  H}_0b^{a,\rm H}_i(T)$ and/or $N\zeta_{\rm  He}x^a_ix^{\rm He}_0b^{a,\rm  He}_i(T)$ to equation (12) if the rate of 
the corresponding reaction ($b^{a,\rm  H}_i$ or $b^{a,\rm He}_i$) exceeded $10^{-14}$ 
cm$^{-3}.$ We also considered the reverse reactions with ${\rm H}^+$ in accordance 
with [44] for the atoms ${\rm N}^0$ and ${\rm O}^0.$

   We will assume that ionizing radiation is generated only behind the shock 
front, and that the generation region has a small optical depth in all lines. 
In this case, we can write for the flux of radiation propagating away from 
(or toward) the star
$$F_l = \frac{h\nu_{1,l}}{2}\int{N_eN^a_i\left(n_1C_{1,l}-n_lC_{l,1}\right)}dz. \eqno(13)$$ 
The expression for the flux in lines with energies below 13.6~eV forming 
behind the front, but outside the transition region, has a similar appearance. 
Below, we will ascribe the subscript $l$ to lines with energies above 13.6 eV 
and use the subscript $k$ in the general case.

In [30], we showed that, in the problem at hand, the temperature of the postshock 
electron gas is appreciably lower than $10^7$~K. At such temperatures, the gas is 
cooled primarily by discrete transitions, but the continuum radiation makes a 
significant contribution to the X-ray emission [36]. Therefore, when calculating 
the spectrum of the cooling zone, we took into account recombination emission by 
H, He, and the C\,VI, C\,VII, O\,VIII, and O\,IX ions [36], as well as free-free 
emission [45]. In contrast to [36], we did not take into account two-photon emission, 
since its contribution is small when $\log N_e > 11$ [46].

If self-ionizing radiation is not generated upstream from the front, the variation of the flux of ionizing photons with distance from the front ($z = 0)$ is described by the law
$$F_l(z)=F_l(0)\exp(-\tau_l),  \eqno(14)$$
with
$$\frac{{\rm d} \tau_l}{\rm d z}=-N
\sum\limits_a{\zeta_a}
\sum\limits_i{x^a_i}
\sum\limits_j{\sigma^a_{i,j,l}}, \eqno(15)$$ 
where the summation is over elements, ions, and shells for 
which $h\nu_l > \chi^a_{i,j}.$ Jumping ahead for a moment, 
we note that the contribution of lines with $\lambda < 912$ {\AA} 
to the cooling function of the preshock gas proves to be insignificant, 
which justifies {\it a posteriori} the use of approximation (14).

The kinetic energy of the directed motion of the gas is converted into 
the thermal energy of atomic nuclei in a layer that we call the shock front. 
According to [32], its thickness is a factor of $(m_p/m_e)^{1/2} \simeq 40$ 
smaller than that of the region following it, in which the temperatures of 
the electrons and heavy particles are equalized. We will treat the shock 
front as a mathematical discontinuity at $z = 0,$ and ascribe the subscript 
$1$ to quantities immediately upstream from the jump and the subscript 2 to 
quantities immediately downstream from the jump. For heavy particles, the 
relationship between the parameters upstream and downstream from the jump is 
obtained by solving the system consisting of equations (5)--(8) and the 
energy conservation law:
$$N_1V_1\left(\frac{\mu_im_{\rm H}V^2_1}{2}+\frac{5kT_{i1}}{2}\right)+\frac{V_1H^2_1}{4\pi}=
N_2V_2\left(\frac{\mu_im_{\rm H}V^2_2}{2}+\frac{5kT_{i2}}{2}\right)+\frac{V_2H^2_2}{4\pi}.$$
Introducing the Alfven velocity $V_a = H/(4\pi\mu_im_{\rm H}N)^{1/2}$ and designating
$$A=\frac{V^2_{A1}}{V^2_1}, \qquad
B = \frac{kT_{i1}}{\mu_im_{\rm H}V^2_1}, \qquad
C = 1+\frac{5A}{2}+5B,$$
we obtain
$$\frac{N_1}{N_2}=\frac{V_2}{V_1}=\frac{1}{8}\left(C+\sqrt{C^2+8A}\right),  \eqno(16a)$$
$$V_{A2}=V_{A1}\sqrt{\frac{V_2}{V_1}},  \eqno(16b)$$
$$T_{i2}=\frac{T_{i1}}{B}\left[\frac{V_2}{V_1}\left(1+\frac{A}{2}+B\right)-\frac{V^2_2}{V^2_1}-\frac{AV_1}{2V_2}\right]. \eqno(16c)$$ 
In the case of a strong shock and a weak magnetic field, i.e., 
when $A,\,B \rightarrow0$, we obtain from (16) the well-known relationships:
$$\frac{N_1}{N_2}=\frac{V_2}{V_1}\simeq\frac{1}{4}, \qquad T_{i2}\simeq\frac{3\mu_im_{\rm H}V^2_1}{16k}. \eqno(16d)$$ 
Recall that, if we use the one-temperature approximation everywhere, 
then, in the strong-shock limit, the temperature immediately behind 
the front will be a factor of two lower than that indicated by formula (16d). 
The electron gas at the viscous jump is compressed adiabatically, and the 
degree of ionization remains virtually unchanged [32]. Hence, we obtain
$$\left(x^a_i\right)_2=\left(x^a_i\right)_1, \qquad
T_{e2}=T_{e1}\left(\frac{V_1}{V_2}\right)^{2/3}.  \eqno(16e)$$

In the strong-shock limit, we have $T_{e2}\simeq2.5T_{e1}.$

\section{Boundary conditions and the solution method}

Before proceeding to a direct analysis for the various flow zones, note that, by adding (6a) and (7), we can obtain with (4) the relationship
{\small
$$j\left[\frac{5}{2}k\left(T_i+x_eT_e\right)+m_{\rm H}\mu_i\frac{V^2}{2}+\frac{H^2}{4\pi N}
-\zeta_{\rm H}x_0^{\rm H}\chi_0^{\rm H}
-\zeta_{\rm He}x_o^{\rm He}\left(\chi_0^{\rm He}+\chi_1^{\rm He}\right)
-\zeta_{\rm He}x_1^{\rm He}\chi_1^{\rm He}
\right]^{z=+\infty}_{z=-\infty}=$$
$$
=-2\sum\limits_{k\neq l}F_k-2\sum\limits_{\rm H,He}F^{rec}\left(h\nu<\chi^{\rm H}_0\right), \eqno(17)
$$
}
expressing the law of conservation of energy for the problem at hand. The term in square brackets is the difference of the energies of the infalling gas far upstream and downstream from the shock, while the right-hand side contains the energy flux carried away by photons with $E < 13.6$~eV. Photons with $E > 13.6$~eV are absorbed upstream from the shock front and in the upper layers of the stellar atmosphere, and are then reprocessed into radiation with $E < 13.6$~eV, which escapes from the system.

The gas temperature $T_0$ far upstream from the front should be of the order of several thousand Kelvin, i.e., $\sim 0.5$~eV. The magnetic energy associated with the infalling plasma is even smaller (see below). On the other hand, for $V_0 = 300$ km\,s$^{-1}$, $E_{kin} \equiv \mu_im_{\rm H}V^2_0/2 \simeq  550$ eV. It therefore follows from (17) that the energy of the infalling material is nearly entirely in the form of kinetic energy, which is spent primarily in radiative cooling, while the losses to gas ionization are small. It is convenient to express the line emission flux $F_k(0)$ in units of $jE_{kin}/2$, introducing the quantities $\delta_k$ characterizing the fraction of energy emitted in the line $k$ downstream from the shock front:
$$F_k(0)=\frac{\mu_im_{\rm H}N_0V_0^3}{4}\delta_k, \eqno(18)$$ 
with $\sum_k\delta_k \le 1.$

\subsection{The preshock region}

In accordance with [26, 32], we expect that the velocity and density of the preshock gas change little, so that
$$V = V_0(1 -\delta V), \qquad N = N_0(1 + \delta N), \qquad H = H_0(1 +\delta H), \eqno(19)$$
where $\delta V,$ $\delta N,$ $\delta H< 1;$ as a consequence of (2), $\delta V =  \delta N,$ and it follows from (5) that $\delta H = \delta N.$

\begin{figure}[h!]
 \begin{center}
  \resizebox{10cm}{!}{\includegraphics{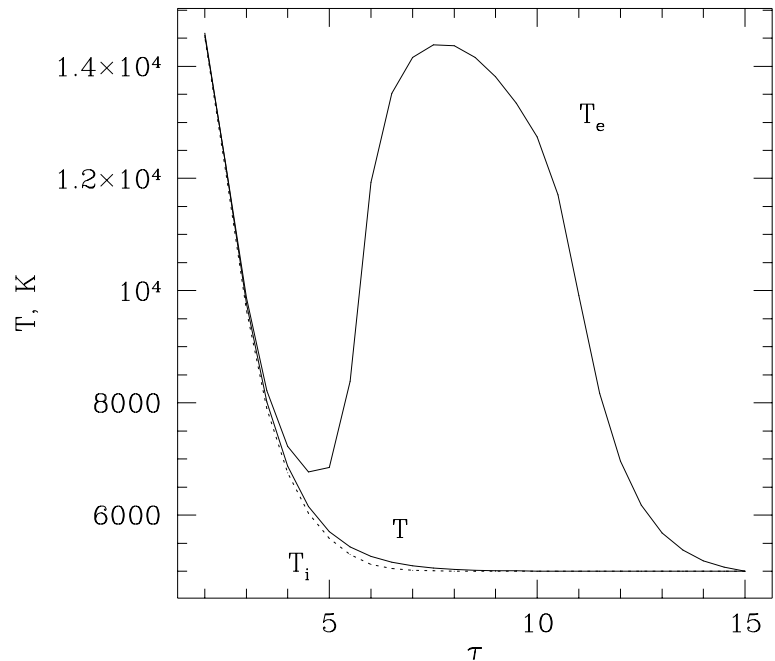}}
  \caption{Electron ($T_e)$ and ion ($T_i)$ temperalures as functions of the optical depth $\tau$ in a hydrogen plasma (see text). The thick line shows the run of the temperature when $T_e = T_i = T.$} 
 \end{center}
\end{figure}


Upstream from the shock front, the electrons are heated directly by photoionization, and we cannot say in advance whether the energy exchange with heavy particles will be efficient enough for their temperature to become essentially the same as that of the electron gas. We can answer this question by solving a model problem in which pure hydrogen is heated by a single ionizing line and the preshock gas has a constant velocity $V= V_0.$

We will provisionally neglect recombination, ionization, and radiative losses due to electron-collisional excitation. We then obtain from equations (6)-(8), (12), and (15):
$$1.5k\left(T_e-T_i\right) = 1.5kT_e(1 + x_e) + \left(h\nu-\chi^{\rm H}_0\right)x_0^{\rm H} + const.$$
Setting $(h\nu - \chi^{\rm H}_0)\simeq1.5kT_e$, and $x^{\rm H}_0\simeq1-x_e$ we have $(T_e- T_i) \sim T_e.$ This means that, as the ionizing photons are absorbed in the stream flowing into the front and the gas is heated, the electron temperature is detached from the ion temperature.

However, when the electron temperature exceeds $10^4$~K, we can no longer neglect cooling of the gas by electron-collisional excitation of the $L_\alpha$ line. Figure 1, which is based on our numerical calculations, shows that this effect limits the growth of $T_e,$ and that the energy exchange with heavy particles rapidly equalizes $T_e$ and $T_i.$ It is important that the temperatures become virtually identical at a rather large optical depth: in the
case shown in Fig.\,1 ($h\nu = 2\chi_0^{\rm H},$ $\delta = 0.1,$ $T_0 = 5000$~K, $V_0 = 300$ km\,s$^{-1}),$ $T_e$ differs from $T_i$ by less than 10\% at $\tau \simeq 5.$ This implies that less than 1\% of the energy pumped into the gas by the flux of ionizing radiation is radiated in the region where the flow has a substantially two-temperature character. Thus, this effect is small, and we will assume that the temperatures of the electrons and heavy particles upstream from the front are equal. The $T(\tau)$ dependence for the model problem in the one-temperature approximation is also shown in Fig.\,1.

Based on the above arguments, we may use the relationships $T_e = T_i = T$ upstream from the shock front, and
$$V\frac{3}{2}k\frac{\rm d}{ \rm d z}\left[T(1+x_e)\right]+VNkT(1+x_e)\frac{\rm d}{ \rm d z}\left(\frac{1}{N}\right)=Q_{ph}-Q_{col}N_e. \eqno(20)$$

If we linearize equation (4) in accordance with (19), we obtain, denoting $\Re = k/\mu_im_{\rm H},$
$$\delta V = \delta N = \delta H \simeq \frac{\Re\left[T(1+x_e)-T_0\right]}{V^2_0}. \eqno(21)$$ 
It is clear from this relationship that the value of $\delta N$ does not exceed 1\% for $T \le 2 \times 10^4$~K. In the same approximation, we can rewrite (20):
$$\frac{3}{2}kV_0\frac{\rm d}{\rm d z}\left[T(1+x_e)\right]=Q_{ph}-Q_{col}N_e. \eqno(22)$$

Heating and cooling of the gas upstream from the front are brought about by photons with energies $h\nu_l > 13.6$~eV $(1 \le l \le N),$ which arrive from the postshock region. To reduce the required computer time, we divided the energy interval from 13.6 to 1000 eV into 50 intervals equal on a logarithmic scale, and treated each interval as a pseudoline with energy $E_l$ equal to the mean value inside the interval and with flux equal to the total flux of the real lines and continuum falling in the given interval. We solved the above equations for these pseudolines, numbered in order of increasing energy.

We took the optical depth $\tau_N$ of the pseudoline with the maximum energy as the independent variable, in place of the $z$ coordinate. Formally, this was done by a termwise division of all differential equations by equation (15b) with $l = N.$ With this choice of independent variable, the equations describing the structure of the preshock region do not depend on $N_0$ and $H_0$ in the case of rarefied gas, and $z(\tau_N)$ linearly depends on $N_0.$ At this stage, we did not consider the change of the preshock gas velocity due to gravity, and used a plane-parallel approximation, assuming that the extent of the preshock H\,II region was small (see below). All else aside, this makes it possible to reduce the number of free parameters. In the case considered of an intermediate-density gas, the structure of the preshock region should depend on $N_0,$ since the level populations of some ions that appreciably contribute to the cooling function (e.g., C\,III, O\,IV, and O\,V) depend on density; however, this dependence proved to be rather weak.

To solve the resulting system of equations, we must fix the value of $V_0,$ which is a parameter, and specify the boundary conditions for $T,$ $z,$ $\tau_l,$ and $x^a_i.$ As stated above, the gas temperature far upstream from the front, i e., at $z \rightarrow -\infty,$ should be of the order of several thousand Kelvin, and we treated $T_0$ as another global parameter of the problem, varying it from 3000 to 6000~K.

The boundary conditions for $z$ and $\tau_l$ $(l < N)$ are obvious: $z = \tau_l = 0$ for $\tau_N = 0.$ The question of the degree of ionization of the material far upstream from the front, i.e., specification of the boundary values for $x^a_i,$ is not so trivial. If we remain in the framework of a plane-parallel geometry, it is natural to assume that the gas is neutral at $z\rightarrow -\infty.$ This automatically implies that the infalling matter is sufficiently abundant to completely absorb the ionizing radiation escaping from beneath the front. Generally speaking, however, this is not the case, since many CTTSs are X-ray sources. It is probable that, owing to the curvature of the magnetic lines of force, the motion of the infalling gas is appreciably non-rectilinear (radial), even at heights $h_X \sim R_*/3$ above the stellar surface. In this case, the hardest part of the postshock radiation will at some time cross the boundaries of the accretion column and will be able to reach the observer, in contrast to lower-energy photons, which are absorbed in the infalling gas at heights $h < h_X.$ (It is probably for this reason that the X-ray spectra of CTTSs resembles the radiation of a plasma with a temperature of $\sim10^7$~K, while the temperature of the post-shock gas is considerably lower.)

Thus, because of the curvature of the lines of force guiding the infall of material, some of the most energetic photons born at the point of intersection of the corresponding line of force with the stellar surface will not reach the gas at heights $h > h_X.$ However, this will to some extent be compensated by the absorption of hard radiation escaping from the limits of the accretion column above other points of the stellar surface. Since there are not currently any observational data about the field geometry in CTTSs, we will assume that, due to this compensation, the plane-parallel approximation remains valid at heights $h > h_X.$ More precisely, we assumed that when $\tau_N \gg 1$, H, He, N, O, and Ne are in atomic form, whereas C, Mg, Si, S, and Fe are singly ionized by Lyman lines of H\,I. Naturally, after solving the equations for this problem, we must verify that the extent of the preshock ionization zone does not exceed the radius $R_1$ of the inner edge of the disk.

In this approach, determining the heat and ionization balance of the preshock region is a boundary-value problem (some of the boundary conditions are specified at the shock front and some at infinity), which we solved as follows. At sufficiently large values of $\tau_N =\tau_N^{max},$ we set trial values of $\tau_l^{max},$ $(l \le N - 1)$ and integrated the equations to $\tau_N = 0.$ If, in this case, some $\tau_l$ differed from zero by more than 0.01, we repeated the integration, changing the values of $\tau_l^{max}$ accordingly.

To solve this problem, we must have a set of values $\delta_l$ for $l \le N,$ which can be found only after determining the structure of the postshock region. However, for this, we must know the temperature and ion composition of the gas immediately upstream from the front. Thus, we obtained a self-consistent solution for the shock-wave structure using an iterative procedure: first, we determined the preshock gas parameters for a trial set of $\delta_l$ values, which made it possible to find first-approximation values of the $\delta_l;$ using these, we again performed the calculations for the preshock region, and so on, until the new $\delta_l$ values were virtually identical to the previous ones.

\subsection{The postshock region}

   Let us now consider the simplifications we used to solve the complete system of equations for the post-shock region. First and foremost, we assumed that, at $T> 10^4$~K, the gas in the recombination region is transparent to the continuum and to its own "cooling"{} radiation; therefore, we did not take photoionization into account in this zone. In addition, we assumed that the density is so high ($N_e > 10^{15}$ cm$^{-3}$ in the transition region between the cooling zone and stellar photosphere, where the "cooling"{}-radiation photons are absorbed, that, due to thermalization, radiative cooling occurs mainly in the continuum at $\lambda>912$ {\AA}, to which all higher-lying layers are transparent. This assumption allows us to completely avoid considering the transition-region structure at this stage, since we are currently interested only in the line emission of "high-temperature"{} ions (see above). The structure and emission of the transition zone is important in analyses of the nature of the so-called "veiling"{} continuum of T Tauri stars and, perhaps, the hydrogen and He\,I emission-line spectra [30]. Note that the gas velocity in the transition region is nearly zero, whereas it is of the order of 300 km\,s$^{-1}$ in the preshock region. Therefore, even if emission in Lyman-series lines carries away appreciable energy from the transition region, this should not strongly influence the thermal balance of the preshock zone.

In the postshock region, it is more convenient to use the variable $T_s = T_ex_e+ T_i$ in place of $T_i.$ An equation for this variable can be derived by adding (6) and (7):
$$V\frac{3}{2}k\frac{{\rm d} T_s}{\rm d z}+VNkT_s\frac{\rm d}{\rm d z}\left(\frac{1}{N}\right)=-Q_{col}N_e.$$

Eliminating the derivative ${\rm d}N^{-1}/{\rm dz}$ from this equation and equation (7), we obtain after some manipulation two equations, which we used in place of equations (6) and (7):
$$\frac{V_2}{N_2}v^2\frac{{\rm d} t_s}{\rm d z}= - \frac{2f_1Q_{col}x_e}{f_2kC_0}, \eqno{23}$$ 
$$\frac{V_2}{N_2}v^2\frac{{\rm d} y}{\rm d z}=\frac{\omega_{ei}x_e-Q_{col}x_e\left(f_2+2C_1y\right)}{1.5kC_0}. \eqno(24) $$
Here, we have introduced the notation $C_0 = \left(T_s\right)_2,$ $C_1 = \Re C_0/V^2_2,$ $C_2 = 0.5 V^2_{A2}/V^2_2,$ $v = V/V_2,$ $t_s = T_s/C_0,$ and $y = T_ex_e/(T_ex_e)_2.$ As in (16), we denote parameters immediately downstream from the viscous jump with the subscript 2; $f_1 = 3v^2 - 2(1 + C_1 + C_2)v + C_1t_s,$ and $f_2 = 3f_1 -2C_1t_s.$ An additional relationship is provided by the algebraic equation
$$v^3 - (1 + C_1 + C_2)v^2 + C_1t_sv+ C_2 = 0,$$
which can be obtained from (2), (4), and (5).

It follows from (16) that, immediately downstream from the viscous jump, the ion temperature reaches $T^i_2> 10^6$~K, while the electron temperature only grows to $\sim 3 \times 10^4$~K. An energy exchange between the electrons and ions begins, and the temperatures are equalized at $T^{max}_e\sim T^i_2/2.$ In this region, up to $T_e = 0.8 T^{max}_e$, we used $T_ex_e$ as the independent variable; for this purpose, we performed a termwise division of all the differential equations by the equation for $y.$ After this, we used $T_s$ as the independent variable, up until the time when the difference between $T_e$ and $T_i$ became less than 1\%. Finally, in the recombination zone, we used a one-temperature approximation and $T_e( 1 + x_e)$ as the independent variable, simultaneously leaving out equation (24) for $T_e.$ We solved the equations in this region up until the time when the gas temperature fell to $T= 10^4$~K.

\section{Discussion}

We solved the above system of radiative gas-dynamical equations numerically, using dedicated software for this purpose. We used a number of test problems to debug this program and verify its correctness, which ultimately gave us confidence in the reliability of the results obtained. In particular, the results of the numerical calculations coincide with analytical solutions we obtained for the case of a hydrogen plasma under certain additional assumptions. Our calculations of the ionization equilibrium for all elements included in the program coincide with the results of [36] in a (stationary) coronal approximation; however, the cooling function $\Lambda(T)$ we obtained for this case differs from that of [36] on average by a factor of two, since we used more accurate formulas for the electron-collisional excitation rates, according to the recommendations of [37].

We used a special subroutine for the numerical solution of systems of stiff differential equations. To test its performance, we computed the ionization balance in a plasma with an arbitrary initial density of ions of various elements for the case of a stepwise change of its temperature: the solutions smoothly converged to the values obtained in the stationary case for a finite temperature. We also calculated the ionization balance and cooling function of an optically thin plasma for the case of isobaric cooling (this situation is similar to the behavior
of the postshock gas [32]) and obtained agreement with [47], with appropriate variation of the atomic data. Finally, we note that, in the computations for the accretion shock wave, the sum of the $x^a_i$ values for each element differed from unity by less than $10^{-6}.$

Let us now turn to the results obtained. First and foremost, it follows from our calculations that the fluxes in the signal lines (see below) are virtually constant if

(1)	we take the initial temperature $T_0$ of the infalling gas far upstream from the front to be in the interval from 3000 to 6000 K;

(2)	in our calculations of the preshock region structure, we do not take into account (pseudo)lines of the ionizing radiation, for which $\delta_l < 10^{-4};$

(3)	 we begin to integrate the equations describing the structure of the preshock region at values $\tau_N^{max}$ exceeding 5 (in accordance with Section 2, we chose the optical depth of the line with the maximum energy of those with $\delta_l > 10^{-4}$ as the independent variable);

(4)	we vary the intensity of the $z$ component of the chaotic magnetic field $H_0$ frozen in the infalling plasma in the range from 0 to 0.1~G.

This last result deserves special attention. The chaotic magnetic field can suppress the electron thermal conductivity if the Larmor radius of an electron in this field is smaller than the mean free path of the electrons between collisions. Using formulas from [32], we can write this condition
$$H\ge0.03\left(\frac{N_e (cm^{-3})}{10^{12}}\right)\left(\frac{10^4}{T(K)}\right)^{3/2}.$$
In the cases of interest for us, $N < 10^{13}$ cm$^{-3}$ and $T\sim 2 \times 10^4$~K in the preshock region (see below). Thus, the assertion in Section 4 {\it a posteriori} indicates the fundamental possibility of determining the structure of an accretion shock wave while simultaneously neglecting the electron thermal conductivity and the chaotic magnetic field frozen in the infalling plasma; this considerably simplifies the system of gas-dynamical equations. It is another question whether the accreted plasma is, in fact, magnetized, and whether we must in reality take into account the heating of the preshock gas by electron thermal conductivity; this question can be answered only after a comparison of the calculated and observed spectra.

Below, we give a relatively detailed description of the shock-wave structure for the case when $V_0 = 300$~km\,s$^{-1}$ and $N_0 = 10^{12}$ cm$^{-3},$ and then show how the main parameters of the wave change as functions of $V_0$ and $N_0.$ We will use 5000~K and 5 for the initial values of $T_0$ and $\tau_N,$ respectively.


\begin{figure}[h!]
 \begin{center}
  \resizebox{10cm}{!}{\includegraphics{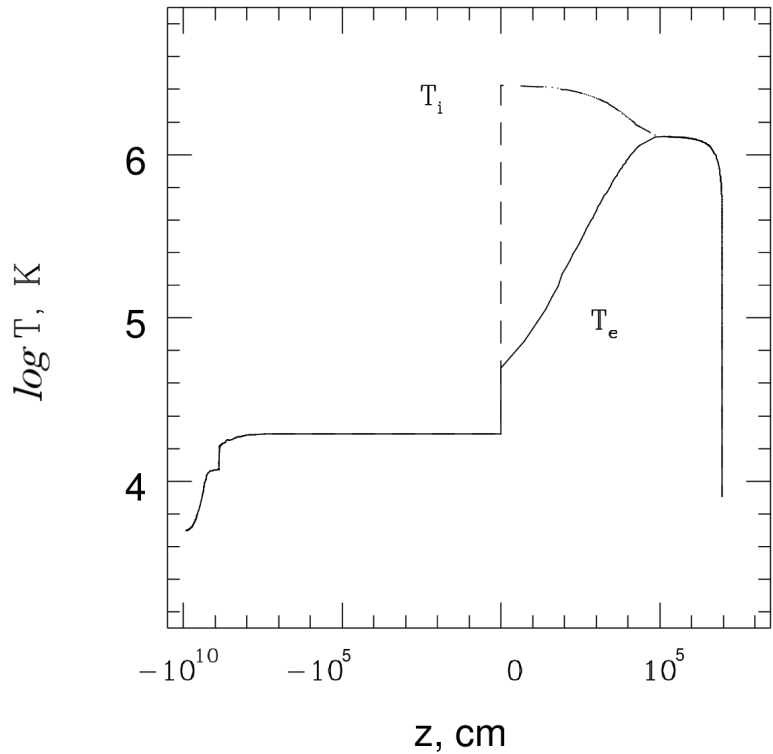}}
  \caption{Variation of electron ($T_e)$ and ion ($T_i)$ temperatures across the shock wave for $V_0 = 300$ km\,s$^{-1}$ and $N_0 = 10^{12}$ cm$^{-3}.$} 
 \end{center}
\end{figure}


Figure 2 shows the distribution of the electron and ton temperatures downstream and upstream from the shock front. We can see that the temperature of the gas approaching the front grows and reaches $T_1 = T(z = 0) \simeq 1.96 \times 10^4$~K immediately upstream from the front. Since $T_1\le2\times 10^4$~ K, in accordance with (21), we can consider our initial assumption that the variations of the preshock gas density and velocity are small to be justified. We will characterize the size of the preshock region by the distance from the front $z_{pre}$ at which the temperature of the infalling gas reaches a value exceeding $T_0$ by 50\%; in the case considered, $z_{pre} \simeq 1.8 \times 10^9$ cm.

In accordance with (16), the ion temperature reaches $T^i_2 \simeq 2.66 \times 10^6$~K immediately downstream from the front, while the electron temperature only grows to $T^e_2\simeq 4.9 \times 10^4$~K. Further, $T_i$ monotonically decreases due to the exchange of energy with the electrons. The postshock electron-gas temperature first
increases to $T^{max}_e \simeq 1.29 \times 10^6$~K at $z_T = z(T^{max}_e) \simeq 1.41 \times 10^5$~cm, then monotonically decreases as the radiative losses begin to dominate over the heating by the ions. At the peak of $T_e,$ the difference of the electron and ion temperatures is less than 1\%, justifying use of a one-temperature approximation. The gas cools to $T = 10^4$~K at a distance $z_{pst} \simeq 9.37 \times 10^6$ cm from the front. Here, we should note that both $z_{pre} \ll R$ and $z_{pst} \ll R_*.$

Since $z_T \ll z_{pst}$ and the width of the region in which the ions are heated is an order of magnitude smaller, our approximation of the front as an infinitesimally thin jump is justified. In addition, $T_e^{max}$ differs from $T_i^{max}/2$ by only 3\%; therefore, this suggests that we can consider the entire postshock region in a one-temperature approximation (see the comment at the end of Section 3).
%
However, there are two reasons why we did not do this. The first is associated with the signal-line profiles: in the one-temperature approximation, their widths immediately downstream from the front due to thermal motion will be systematically larger than observed. Second, in the region where the ion temperature exceeds $T_i^{max}/2,$ nuclear reactions converting deuterium to $^3$He can occur. We plan to treat the corresponding set of problems in a separate study, and here restrict our consideration to the comment that, according to our
estimates, the energy release of these nuclear reactions is sufficiently small that they should exert virtually no effect on the shock-wave structure.

Figure 3 shows the spectra of the ionizing radiation escaping from the gas cooling zone for three values of $V_0$ and for $\log N_0 = 11,$ or, more precisely, the values of $\delta_l$ for the 50 energy intervals (pseudolines). The dashed line at $\delta = 10^{-4}$ enables us to determine which pseudolines play an appreciable role in the heat and ionization balance in the preshock region. In particular, it follows from this figure that, for $V_0 = 300$ km\,s$^{-1}$, the line with the maximum energy is that with $N = 44$ ($E_{44} \simeq 571$~eV), whose optical depth we used as the independent variable in this case.

%
\begin{figure}[h!]
 \begin{center}
  \resizebox{6cm}{!}{\includegraphics{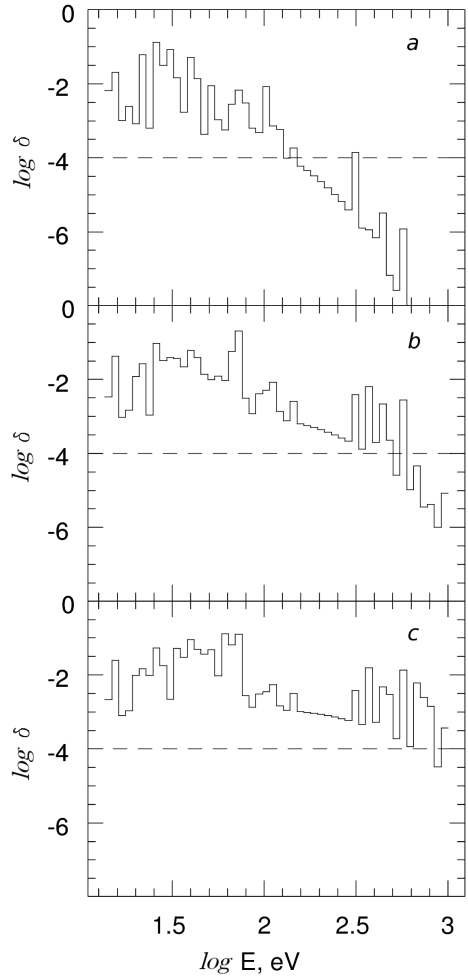}}
  \caption{Emission spectrum generated behind the shock front for three values of the preshock gas velocity: $V_0 = 200$ (a), 300 (b), and 400 km\,s$^{-1}$ (c). In all cases, $\log N_0 =11.$} 
 \end{center}
\end{figure}

We will now consider how the gas ionization varies along the flow. The preshock H\,II region is approximately half the size of $z_{pre}:$ $z_{\rm H\,II}\simeq 7 \times 10^8$~cm. As the gas approaches the front, the degree of ionization of the hydrogen steadily increases, reaching $x^{\rm H}_0 = 7 \times 10^{-4}.$ Immediately upstream from the front, 96\% of the helium is in the form of He\,III; carbon and oxygen ions up to a charge of $+4,$ nitrogen, neon, magnesium, and silicon ions up to a charge of $+5,$ and sulfur and iron ions up to a charge of $+6$ all have abundances exceeding 1\%.

Immediately downstream from the front, the time scale for heating of the electrons is much shorter than the time scale for ionization and recombination; therefore, the ionization equilibrium cannot be established after the rapid growth of the temperature. As a result, the abundance of, for example, C\,IV ions exceeds 1\% up to $T_e \simeq 750\,000$~K, and that of O\,VI ions becomes less than 1\% only at $T_e \ge 1.1 \times 10^6$~K. Recall that, in an equilibrium situation, these ions are virtually absent at temperatures exceeding $2\times 10^5$ and $6\times 10^5$~K, respectively. In short, the degree of ionization of the gas is considerably lower than the equilibrium value over nearly the entire region of growth of $T_e.$

As the ion and electron temperatures equalize, the time scale for the growth of $T_e$ decreases. Therefore, near the peak of the electron temperature, the ion abundances roughly correspond to the equilibrium values for $T^{max}_e.$ However, soon after the electrons have begun to cool, the cooling time scale becomes much shorter than the recombination time scale, and the equilibrium again breaks down. Now the degree of ionization is higher than the equilibrium value: for instance, the C\,IV and O\,VI ions reach their maximum abundances only at temperatures of the order of $2 \times 10^4$ and $8 \times 10^4$~K, respectively!

It follows from the above discussion, in particular, that the recombination of ions with charges less than six occurs at temperatures much lower than in an equilibrium situation. However, this means that, in our case, the role of dielectronic recombination is less important than usual. This suggests that, even if our description of dielectronic recombination is not entirely correct, this does not strongly affect the results obtained.

We calculated the shock-wave structure in the approximation of volume radiative-energy losses; now we must test to what extent this is justified. We will start from the fact that typical values of the critical density for allowed dipole transitions are $N_{cr} = A_{k1}/C_{k1} \sim 10^{17}-10^{18}$ cm$^{-3}.$ For the part of the preshock H\,II region where $N_e \sim N_0,$ the optical depth in the $L_{\alpha}$ line proved to be $10^6$ (in the model with $V_0 = 300$ km\,s$^{-1}$ and $N_0 = 10^{12}$ cm$^{-3}).$ Therefore, the probability of escape of a $L_{\alpha}$ photon from this region, which is to order of magnitude equal to $\tau^{-1/2}\sim10^{-3},$ considerably exceeds the probability of thermalization, equal to $N_e/N_{cr}\sim 10^{-6}.$ Beyond the region considered, the hydrogen becomes neutral, and the optical depth of the $L_{\alpha}$ line grows rapidly, reaching $\sim10^8$ at $z = z_{pre}$; however, the electron number density and gas temperature simultaneously decrease. This means that we can use the volume energy-loss approximation throughout the region where the contribution of the $L_{\alpha}$ line to the cooling function plays an appreciable role. For resonant lines of other elements with $\lambda > 912$ \AA, our assumption of the volume character of the preshock radiative losses is even better satisfied; for example, for the $\lambda$1550 \AA line of C\,IV, $\tau \simeq 6 \times 10^3.$

Similar estimates based on the results of our calculations show that the volume energy-loss approximation is also justified downstream from the shock front. In this very region, the gas turned out to be transparent to the Lyman continuum, confirming that we can neglect photoionization processes here.

We have intentionally distinguished the postshock region with $T> 10^4$~K. The reason is that, at lower temperatures, H\,II and He\,II, which should efficiently absorb photons with $E > 13.6$~eV traveling toward the star, begin to intensely recombine. In this transition zone between the shock wave and the stellar atmosphere, the gas velocity is nearly zero, and the gas density is several orders of magnitude higher than upstream from the front: for example, in the model considered, $V \simeq 0.9$ km\,s$^{-1}$ and $N \simeq 4 \times 10^{14}$ cm$^{-3}$ for $T = 10^4$~K. This indicates that we cannot use the volumeloss approximation for $T > 10^4$~K; instead, we must solve for the heating of the stellar atmosphere by external short-wavelength radiation taking into account the effects of radiative transfer in the lines and continuum, as was done in [48].

The calculations show that up to the time when our approach is no longer adequate, the abundances of ions with charges exceeding $+1 $ become negligible for all the elements we have considered. It is likely that ions with charges $+2$ and higher will not appear in significant numbers in the transition zone, due to the high density of material there. This statement is based on the calculation of the transition-region structure carried out using the method we applied for the preshock zone, i.e., without taking into account the effects of thermalization of the radiation. The true temperature distribution inside the heating region may considerably differ from the calculated distribution, but we do not see why the efficiency of radiative recombination and charge exchange -- which give rise to the near complete absence of highly-charged ions in our model of the transition zone -- should decrease.

Table 1 shows the main parameters characterizing the structure of an accretion shock wave as functions of $V_0$ and $N_0.$ We can see that the preshock gas temperature decreases somewhat with increasing $N_0$ as a consequence of the reduced contribution of intercombination lines to the cooling function. This effect influences the geometric factors to a much lesser extent: $z_{pre}$ and $z_{pst}$ are nearly inversely proportional to $N_0.$ The dependence of the parameters on $V_0$ is rather trivial, and, in our view, does not require additional discussion.

\medskip
\begin{table}
  \caption{Parameters of the structure of an accretion shock wave}
 \label{table1}
\begin{center}
\begin{tabular}{|c|c|c|c|c|}
\hline
${\rm V_0},$ km\,s$^{-1}$ & 200 & \multicolumn{2}{|c|} {300} & 400 \\ 
\hline
${\rm \log\, N_0,}$ cm${}^{-3}$& 12 & 11 & 12 & 11 \\ 
\hline
${\rm z_{pre}},$ cm & $1.3\times10^8$ & $2.8\times10^{10}$ & $1.8\times10^9$ & 
$8.4\times10^{10}$  \\
${ z_{\rm H\,II},}$ cm & $1.3\times10^8$ & $2.1\times10^{10}$  & $7.0\times10^8$ &
$2.4\times10^{10}$   \\
${ x({\rm H\,I})}$     & 0.0027  & 0.0008  &  0.0007  & 0.0005  \\
${ x({\rm C\,IV})}$    & 0.096 & 0.17 & 0.16  & 0.093 \\
${ \tau_{{\rm C\,IV}\, 1548}^{pre}}$  & $2.2\times 10^2$ & $5.6\times 10^3$ & $5.9\times 10^3$ 
& $2.0\times 10^4$  \\
${ T_1},$ K     & $1.7\times10^4$ & $1.8\times10^4$ & $2.0\times10^4$ &
$1.8\times10^4$   \\
${ T_2^e},$ K     & $4.2\times10^4$ & $4.5\times10^4$ & $4.9\times10^4$ & 
$4.6\times10^4$   \\
${ T_2^i},$ K     & $1.2\times10^6$ & $2.7\times10^6$ & $2.7\times10^6$ & 
$4.7\times10^6$  \\
${ T_e^{max}},$ K     & $5.6\times10^5$ & $1.3\times10^6$ & $1.3\times10^6$ & 
$2.3\times10^6$  \\
${ z(T_{max})},$ cm     & $2.2\times10^4$ & $1.4\times10^6$ & $1.4\times10^5$ & 
$3.2\times10^6$  \\
${ z_{pst}},$ cm & $1.4\times10^6$ & $9.3\times10^7$ & $9.4\times10^6$ & 
$2.0\times10^8$  \\
\hline
\end{tabular}
\end{center}
\end{table}

Table 2 lists the line fluxes for some ions, normalized to the total flux of the $\lambda\lambda$ 1548, 1551 doublet of C\,IV and expressed in percent. For determining the absolute flux values using relationship (18), the values of $\delta$[C\,IV 1550] are also indicated. Note that, in all our models, as in all T Tauri stars, the $\lambda$1550 doublet of C\,IV proved to be the most intense line at wavelengths 1200--2000 \AA, apart from the $L_{\alpha}$ $\lambda$1215~{\AA} line, of course.

Some ions have appreciable abundances in three regions: in the preshock H\,II region, immediately downstream from the front (where the electron temperature grows), and, finally, in the recombination zone.

Therefore, generally speaking, these lines should have three-peaked profiles, due to the different gas velocities in these regions: $V_0,$ $V_0/4,$ and $\simeq0.$ Note that multicomponent structure is, indeed, observed in some lines of highly-charged ions in the spectra of T Tauri stars [24]. Our calculations indicate that the relative intensities of these peaks change considerably depending on the values of $V_0$ and $N_0,$ and that this dependence is different for different ions. For instance, if $\log N_0 = 12,$ the ratio of the fluxes of the C\,IV $\lambda$1550 A doublet from the post-shock and preshock regions is 1 : 17 for $V_0 = 200$ km\,s$^{-1}$ and 2 : 1 for $V_0 = 300$ km\,s$^{-1}.$ The flux from the post-shock region for the intercombination C\,III $\lambda$1908 {\AA} line is negligible in all cases considered, while for the O\,VI $\lambda$1031 \AA{} doublet, on the contrary, the intensity of emission from the preshock region is negligible.

\medskip

\begin{table}
  \caption{Fluxes in lines of some ions (percent relative to the total flux in the C\,IV $\lambda\lambda$1548, 1553 doublet)}
 \label{table2}
\begin{center}
\begin{tabular}{|c|c|c|c|c|}
\hline
${ V_0},$ km\,s$^{-1}$ & 200 & \multicolumn{2}{|c|} {300} & 400 \\ 
\hline
${ \log\, N_0,}$ cm${}^{-3}$& 12 & 11 & 12 & 11 \\ 
\hline
${ \delta ({\rm C\,IV}}$ 1550) & 0.045 & 0.042 & 0.048 & 0.049  \\
\hline
${\rm O\,VI}$ 1032+1038 \AA & 900 & 440 & 330 & 220  \\
${\rm N\,V}$ 1239+1243 \AA & 54 & 29 & 22 & 19  \\
${\rm Si\, IV}$ 1394+1403 \AA & 16 & 21 & 26 & 12  \\
${\rm O\, III]}$ 1661+1666 \AA & 10 & 92 & 9.1 & 69  \\
${\rm Si\, III]}$ 1892 \AA & 2.2 & 3.5 & 0.9 & 1.3  \\
${\rm C\, III]}$ 1908 \AA & 0.2 & 1.6 & 0.2 & 0.8  \\
\hline
\end{tabular}
\end{center}
\end{table}

As we noted above, there is no thermalization of the signal-line emission. Since the parameters of the preshock H II regions obtained in our numerical calculations are nearly identical to those used in [30], we suggest that, if the preshock gas density does not strongly exceed $10^{12}$ cm$^{-3},$ radiation at $\lambda > 912$ {\AA} leaves the shock nearly unimpeded. This means that, in spite of the considerable optical depth, the intensity ratio for the components of the C\,IV $\lambda$1550 doublet should be 2 : 1, since this is equal to the ratio of the electron-collisional excitation coefficients for the corresponding levels, which is nearly equal to the ratio of the statistical weights of these levels. The observed intensity ratio for the $\lambda$1548 and $\lambda$1551 lines in T Tauri stars is 2 : 1, but, in accordance with the arguments above, this should in no way be taken as proof that these lines are optically thin, as asserted in [3, 7, 24].

In the signal lines, whose optical depth considerably exceeds 1, the observed emission flux should depend on the area of the accretion zone and its orientation with respect to the observer, while the fluxes of optically thin lines are determined by the emission measure. Therefore, the relative intensities of the signal lines in the spectrum of a star, generally speaking, should differ from those listed in Table 2. For the same reason, in stars with different accretion-zone geometries, both the line profiles and the overall appearance of the emission-line spectrum should differ, even if their values of $V_0$ and $N_0$ are identical.
 
\section{Conclusion}

Based on the discussion in the previous section, we expect that, in the framework of our assumptions, we have correctly calculated the line-emission intensities for ions with charges above $+1.$ We have also shown that the results of our numerical calculations confirm {\it a posteriori} the validity of our main assumptions, adopted to simplify the system of hydrodynamical equations used to model the structure of an accretion shock wave. It is important that the relative intensities and profiles of the spectral lines strongly depend on $V_0$ and $N_0;$ consequently, the results obtained have diagnostic value. All this leads us to believe that comparison of the calculated fluxes and profiles of the C\,III, C\,IV, Si\,III, Si\,IV, etc., lines with those observed in the spectra of young stars may provide a base for elucidating the origin of the activity of these objects.

\bigskip

{\it Acknowlegements}

The author is grateful to L.\,A. Vainshtein, M.\,A. Livshits, and A.\,F. Kholtygin for useful discussions, and to M. Pavlov, I.\,I. Antokhin, and M.\,E. Prokhorov for advice concerning the computer software. Special thanks are due to K.\,V. Bychkov, who taught me a lot. This work was supported by the Russian Foundation for Basic Research (project code 96-02-19182).

\bigskip

{\bf \large{ References }}
\medskip

1.	Appenzeller, I. and Mundt, R., Astron. Astrophys. Rev., 1989, vol. 1, p. 291.

2. 	Bertout, C., Ann. Rev. Astron. Astrophys., 1989, vol. 27, p. 351.

3.	 Imhoff, C. and Appenzeller, I., {\it Exploring the Universe with the IUE Satellite }, Kondo, Y. el al., Eds., Boston: Kluwer, 1987, p. 295.

4.	 Lamzin, S.A., Cand. Sci. Dissertation, Moscow: Space Research Institute, 1985.

5.	 Kuhi, L.V., Astrophys. J., 1964, vol. 140, p. 1409.

6.	 Bisnovatyi-Kogan, G.S. and Lamzin, S.A., Astron. Zh., 1977, vol. 54, p. 1268.

7.	Lago, M.T.V.T., Mon. Not. R. Astron. Soc., 1984, vol.210, p. 323.

8.	Hartmann, L., Edwards, S., and Avrett, E., Astrophys. J., 1982, vol. 261, p. 279.

9.	De Campli W.M., Astrvphys. J., 1981, vol. 244, p. 124.

10.	Lynden-Bell, D. and Pringle, J.E., Mon. Not. R. Astron. Soc., 1974, vol. 168, p. 603.

11.	Bertout, C., Basri, G., and Bouvier, J., Astrophys. J., 1988, vol.330, p. 350.

12.	Camenzind, M., Rev. Mod. Astron., 1990, vol. 3, p. 234.

13.	Giovannelli, F., Rossi, C., Errico, L., Vittone, A., et al., NATO ASl Series, 1990, vol. 340, p. 97.

14.	Konigl, A., Astrophys. J., 1991, vol. 370, p. L79.

15.	Konigl, A. and Ruden, S.P., {\it Protostars and Planets III}, Levy, E. and Lunine, J., Eds., Tucson: Univ. Arizona Press, 1992, p. 641.

16.	Grinin, V.P. and Mitskevich, A.S., {\it Flare Stars in Star Clusters}, Mirzoyan, L.V. el al., Eds., Dordrecht: Reidel, 1990, p. 343.

17.	Hartmann, L., Hewett, R., and Calvet, N., Astrophys. J., 1994, vol. 426, p. 669.

18.	Solf, J. and Bohm, K.-H., Astrophys. J., 1993, vol. 410, p. L31.

19.	B{\"o}hm, K.-H. and Solf, J., Astrophys. J., 1994, vol. 430, p. 277.

20.	Lamzin, S.A., Astron. Zh., 1989, vol. 66, p. 1330.

21.	Lamzin, S.A., Proc. IAU Coll. 129, Bertout, C. et al., Eds., Paris, 1991, p. 461.

22.	Krautter, J., Appenzeller, I., and Jankovics, I., Astron. Astrophys., 1990, vol. 236, p. 416.

23.	Hamann, F. and Persson, S.E., Astrophys. J., Suppl. Sen, 1992, vol. 82, p. 247.

24.	Gomez de Castro, A.I., Lamzin, S.A., and Shatskii, N.I., Astron. Zh., 1994, vol. 71, p. 609.

25.	Guenther, E., PhD Dissertation, MPIA, Heidelberg, Germany, 1993.

26.	Kaplan, S.A. and Pikel'ner, S.B., {\it Fizika mezhzvezdnoi sredv (Physics of the Interstellar Medium)}, Moscow: Nauka, 1963.

27.	Kaplan, S.A. and Pikel'ner, S.B.,{\it Mezhzvezdnaya sreda (The Interstellar Medium)}, Moscow: Nauka, 1979.

28 Klimishin, I.A., {\it Udarnye volny v obolochkakh zyezd (Shock Waves in Stellar Envelopes)}, Moscow: Nauka, 1984.

29. Lipunov, V.M., {\it Astrofizika neitronnykh zvezd (Astrophysics of Neutron Stars)}, Moscow: Nauka, 1989.
30.	Lamzin, S.A., Astron. Astrophys., 1995, vol. 295, p. L20.

31.	Allen, C.W., {\it Astrophysical Quantities}. 3rd Ed., London: Athlone, 1973.

32.	Zel'dovich, Ya.B. and Raizer, Yu.P.,{\it  Fizika udarnykh voln i vysokotemperaturnykh yavlenii (Physics of Shock Waves and High-Temperature Phenomena)}, Moscow: Nauka, 1966.

33.	Landau, L.D. and Livshits, E.M., {\it Elektrodinamika sploshnykh sred (Electrodynamics of Continuous Media)}, Moscow: Nauka, 1992.

34.	Verner, D.A. and Yakovlev, D.G., Astron. Astrophys., Suppl. Ser., 1995, vol. 109, p. 125.

35.	Hummer, D.G., Mon. Not. R. Astron. Soc., 1994, vol. 268, p. 109.

36.	Landini, M., Monsignori Fossi B.C., Astron. Astrophys., Suppl. Ser., 1990, vol. 82, p. 229.

37.	Atomic Data and Nuclear Data Tables, 1994, vol. 57, p. 1.

38.	Burgess, A. and Summers, H.P., Astrophys. J., 1969, vol. 157, p. 1007.

39.	Reisenfeld, D.B., Raymond, J., Young, A.R., and Kohl, J.L., Astrophys. J., 1992, vol. 389, p. L37.

40.	Beigman, I.L., Vainshtein, L.A., and Chichkov, B.N., Zh. Eksp. Teor. Fiz., 1981, vol. 80, p. 964.

41.	Borovskii, A.V., Zapryagaev, S.A., Zatsarinnyi, O.I., and Manakov, N.L., Plaznia mnogozaryadnykh ionov (A Plasma of Multiply-Charged Ions), St. Petersburg: Khimiya, 1995.

42.	Kunc, J.A. and Soon, W.H., Astrophys. J., 1992, vol. 396, p. 364.

43.	Bates, D.R., {\it Atomic and Molecular Processes}, New York: Academic, 1962.

44.	Butler, S.E. and Dalgarno, A., Astrophys. J., 1980, vol. 241, p. 838.

45.	Hummer, D.G., Astrophys. J., 1988, vol. 327, p. 477.

46.	Drake, S. and Ulrich, R., Astrophys. J., 1981, vol. 248, p. 380.

47.	Schmutzler, T. and Tscharnuter, W., Astron. Astrophys., 1993, vol. 273, p. 318.

48.	Sakhibullin, N.A. and Shimanskii, V.V., Astron. Zh., 1996, vol. 73, p. 73.

\end{document}